\documentclass[dblblindworkshop,final]{article}
\usepackage{neurips_2026}

\usepackage{lipsum}

\usepackage{multirow}
\usepackage{subcaption}
\usepackage{url}

\usepackage{booktabs,tabularx}
\usepackage{ragged2e}
\usepackage{tikz}
\usepackage{graphicx}

\usepackage{listings}
\usepackage{xcolor}
\usepackage{enumitem}
\usepackage{hyperref}
\usepackage{siunitx}
\usepackage{pdflscape}
\usepackage{rotating}
\usepackage{float}
\usepackage{dblfloatfix}

\usepackage[linesnumbered]{algorithm2e}
\SetAlCapFnt{\footnotesize}
\SetAlCapNameFnt{\footnotesize}
\SetAlCapSkip{4pt}

\usepackage{localmacros}

\begin{document}

\date{}

\title{How Agents Ask for Permission: User Permissions for AI Agents, from Interfaces to Enforcement}

\workshoptitle{The Third Workshop on Agents in the Wild: Safety, Security, and Beyond}

\author{
{Alexandra E. Michael}\\
University of Washington\\
\texttt{aemichae@cs.washington.edu}
\And
{Franziska Roesner}\\
University of Washington\\
\texttt{franzi@cs.washington.edu}
} %

\begin{abstract}

As AI agents grow in prevalence, so too do their risks to users, creating a need for robust and usable systems for agent permissions.
Many recent proposals implement \emph{product-level} permissions, where the same security policy is applied for all users of a given agent framework.
Yet different users have varying needs, necessitating support for \emph{user-level} permission policies in agentic AI systems.

We survey 21 proposed user-level agent permissions systems and construct a taxonomy of approaches to specifying, deriving, and enforcing user-level permissions.
We further analyze and compare against five commercial agents. %
We find that despite a wide diversity of proposals, sizable gaps remain in usability, transparency, and user control, particularly between academic and commercial agents.

\end{abstract}

\maketitle

\section{Introduction}

AI agents that combine LLMs with executable tools to take on autonomous tasks are increasingly prevalent.
While agents are powerful, they also pose serious dangers~\cite{kim-et-al:agent-sec-sok:usenix26,krebs:instagram-bot-hack:2026,metr:hugging-face:2026,olson:openai-australia-breach:axios2026,chia-and-cress:claude-hack:bbcjul2026,bhuiyan:muse-home-address:guardian2026,bagdasarian-et-al:contextual-agent-sec:2026}.
Agents amplify all the existing risks of LLMs---from prompt injection and jailbreaking, to disclosing sensitive information~\cite{owasp:llm-top-10:2024,das-et-al:llm-security-survey:2025}---with the ability to act directly on files, over network connections, and more.

Secure agentic systems address this challenge by restricting an agent's actions according to a security policy.
Often, policies are defined at a global or \emph{product} level, and applied uniformly across all users.
Yet what makes for ``correct'' security policies, or permissions, depends on the user and context~\cite{barth-et-al:contextual-integrity:2006,benthall-et-al:contextual-integrity-in-cs:2017,bagdasarian-et-al:contextual-agent-sec:2026}.
For example, one product-level security policy might block passport information from being sent to any websites other than recognized airlines.
Such a policy does not account for users who want their information shared with a specialty travel booking service, nor for those who never want it shared at all.

The context-dependent nature of permissions drives a need for systems that support not just product-level, but \emph{user-level} agent security policies.
In this work, we explore, first, the current state of the art on handling user-level permissions in agentic AI systems; and second, where the agentic security community should go from here.
We approach these topics via three research questions:

\textbf{RQ1:} How do proposed agent permissions systems allow users to specify their policies?
\textbf{RQ2:} How are user-level policies \emph{internally} specified, derived from user-selected options, and enforced?
How much user transparency and control is there at each step?
Finally, \textbf{RQ3:} What gaps remain in the existing work on user-level permissions systems for AI agents, both \textbf{(a)}~within the literature and \textbf{(b)}~between the literature and commercial agents?

We address these questions by reviewing 21 secure agentic systems that implement user-level permissions as of the early summer of 2026.
From this survey, we construct a taxonomy of approaches in agentic systems to \emph{specifying} permissions policies, both at the user interface and internally; \emph{deriving} internal policies from user input; and \emph{enforcing} polices at run time.
We further analyze five prominent commercial AI agents to understand how agent permissions systems deployed in practice compare to those in the literature.

We identify several cross-cutting concerns as themes in the literature, including minimal user overhead, formalized policy specifications, and deterministic enforcement.
We also find substantial gaps: none of the proposals we review fully address all three concerns, though a few come close.
On the other hand, the commercial agents we evaluate often force users to choose between high-overhead user-in-the-loop permissions, and LLM auto-reviewers with little transparency or user control.
Commercial agents do, however, allow users to modify policies over time, an option largely overlooked in the literature.

Our contributions:

\begin{itemize}
    \item A survey of proposals for handling user-level permissions in the agentic security literature.
    \item A taxonomy of approaches to specifying, deriving from user input, and enforcing user-level permissions policies in AI agents.
    \item A comparative analysis of user permissions handling in five commercial AI agents.
    \item Important gaps and areas for future work on user-level permissions in agentic systems.
\end{itemize}

\section{Background}
\label{sec:background}

\subsection{AI Agents}

In line with related work~\cite{kim-et-al:agent-sec-sok:usenix26}, we use ``AI agent'' and ``agentic AI systems'' interchangeably, meaning systems that use LLMs in combination with non-AI components to call and execute external tools~\cite{debenedetti-et-al:camel:satml26,wu-et-al:isolategpt:2025,masterman-et-al:agent-architectures-survey:2024}.
We focus on permissions for the agentic framework as a whole.
Approaches targeting standalone LLMs only~\cite{microsoft:aici:2026,rebedea-et-al:nemo:emnlp23}, while complementary, are out of scope.

\subsection{User-Facing Permissions}

User-facing permissions have a long history in traditional, non-AI systems.
Established approaches include access control lists (ACLs) and capabilities, built on principles like \emph{least privilege} and \emph{failing closed}~\cite{saltzer-and-schroeder:security-principles:1975,smith:contemporary-security-principles:2012}.
All user permission systems face the challenge of balancing security with usability: How easily can users specify policies that match their security needs?

This problem is hard in part because appropriate permissions depend on the user and context~\cite{barth-et-al:contextual-integrity:2006,benthall-et-al:contextual-integrity-in-cs:2017}.
Asking users to make every policy decision manually is risky, as users often over-grant permissions when overwhelmed by prompts~\cite{choi-et-al:privacy-fatigue:2018,keith-et-al:privacy-fatigue:2014,vance-et-al:fog-of-warnings:2019}.
On the other hand, automatically granting permissions minimizes prompts at the cost of reduced transparency and user control~\cite{felt-et-al:how-to-ask:2012}.
AI agents present even greater difficulty.
Agents are inherently nondeterministic, possess an arbitrary range of actions, and unavoidably conflate instructions and data~\cite{owasp:llm-top-10:2024,das-et-al:llm-security-survey:2025,bagdasarian-et-al:contextual-agent-sec:2026}.
These challenges make the security-usability tradeoff more acute than ever, amplified by pressure to preserve the agent's run-time utility.

\subsection{Agentic Security}

Recent literature explores a range of approaches to securing AI agents~\cite{kim-et-al:agent-sec-sok:usenix26,christodorescu-et-al:systems-sec-for-agents:2026}, many of which form the basis of user-focused agent permission systems.
AI guardrails directly constrain the output of a decision-making LLM by using a secondary model to identify policy violations~\cite{microsoft:aici:2026,rebedea-et-al:nemo:emnlp23,invariantlabs-ai:invariant:2026}.
Multi-LLM systems, such as CaMeL, create a separation of privilege by designating distinct LLMs for different tasks~\cite{debenedetti-et-al:camel:satml26,wu-et-al:isolategpt:2025}.
Other approaches apply established security mechanisms to agent frameworks, such as mobile-style permission policies~\cite{wu-et-al:isolategpt:2025,wu-et-al:automating-data-access:arxiv25} or information flow control~\cite{wu-et-al:f-secure:2024,debenedetti-et-al:camel:satml26,li-et-al:ace:ndss26,costa-et-al:fides:2025,meijer:guardians:acm26,amin:guardians:2026}.

No existing solution is perfect.
AI guardrails are vulnerable to adaptive attacks~\cite{nasr-et-al:attacker-moves-second:arxiv25}, privilege separation struggles to maintain run-time flexibility~\cite{debenedetti-et-al:camel:satml26}, and deterministic systems demand exacting policy specifications. %
Furthermore, much of the literature focuses on product- rather than user-level permissions.
For instance, both CaMeL and \textsc{Fides} fix security policies at the architectural level~\cite{debenedetti-et-al:camel:satml26,costa-et-al:fides:2025}.
While \textsc{Prudentia}~\cite{kolluri-et-al:prudentia:2026} later built on \textsc{Fides} to allow user influence over run-time policies, 
CaMeL has yet to be extended for user-specific permissions~\cite{debenedetti-et-al:camel:satml26}.

\section{Methods}
\label{sec:methods}

\subsection{Literature Review and Taxonomization}
\label{ssec:lit-review}

We aim to build a broad, high-level taxonomy of user-level agent permission systems.
We therefore focused our literature search on collecting a representative set of recent and influential works.
In total, we reviewed 21 systems and proposals from 2024 -- 2026, including 8 peer-reviewed publications, 12 pre-prints, and one open-source system without an associated paper~\cite{provos:iron-curtain:2026}.\footnote{Several of the peer-reviewed works were reviewed as pre-prints and have since been published. In addition, two of the pre-prints enumerated here, PCAS and FORGE, are iterations of the same work~\cite{palumbo-et-al:policy-compiler:arxiv26,palumbo-et-al:forge:2026}. We reviewed PCAS prior to FORGE's release, and retained it in our analysis due to the substantial architectural and design differences between the two.}
We began by identifying influential and/or recent agentic security papers, e.g.,~\cite{wang-et-al:agentspec:icse26,shi-et-al:progent:arxiv25,palumbo-et-al:policy-compiler:arxiv26,debenedetti-et-al:camel:satml26}, using Google Scholar and existing broad surveys~\cite{kim-et-al:agent-sec-sok:usenix26,das-et-al:llm-security-survey:2025} as points of reference to avoid missing prominent or representative prior work. 
We then recursively followed citation paths (``snowballed'') to incorporate those papers' related works.
We repeated this process multiple times over the course of the literature review, conducted from April -- early June 2026, as new related works were released.

These steps produced an initial pool of 30 relevant papers and proposals.
We then conducted a preliminary review.
This review consisted of lightly reading each paper to ensure it was in scope, namely by (1) targeting tool-calling AI agents; and (2) either proposing or implementing a concrete framework, mechanism, or technique for user-level permission policies.
Preliminary reviews eliminated 9 works that either targeted standalone LLMs over agents, only implemented \emph{product-} rather than \emph{user-}level permission policies, or did not describe any concrete system.

In parallel with preliminary reviews, we sketched an initial user-level permissions taxonomy covering (a) policy specifications, (b) derivation of internal specifications from user selections, and (c) enforcement mechanisms.
We based the initial taxonomy on a combination of established security principles and initial reviews of a few representative papers.
The taxonomy was then iteratively refined and deepened over the course of the full literature review.

We read each of the 21 works in depth and took notes on each system's threat model and approach to specifying, deriving, and enforcing permission policies, before labeling it in accordance with the taxonomy.
In cases where the initial taxonomy sketch was insufficient to represent a given work, we deepened the taxonomy accordingly, then revisited previously labeled works to ensure their labels remained consistent.
These updates primarily occurred early in the literature review, as the taxonomy quickly converged.
Finally, after all works had been assigned labels, we validated all assignments with a final review pass.
The taxonomy development and paper categorization process was undertaken primarily by the first author, with periodic feedback from the last author.

\subsection{Commercial Agent Walkthroughs}
\label{ssec:agent-analysis}

We evaluate five prominent commercial agents to examine how user permissions for agents in active use differ from proposed systems.
We consider agents across two companies and three settings: OAuth-enabled chatbots (Claude~\cite{claude}, ChatGPT~\cite{chatgpt}), desktop agents (Claude Cowork~\cite{claude-cowork}, Codex~\cite{codex}), and an agentic browser (ChatGPT agent mode~\cite{chatgpt}).
For each agent, we conducted a walkthrough~\cite{light-et-al:walkthrough-method:2018} exploring how users specify permissions, and how (reliably and consistently) user-specified policies are applied in practice.
For details, see Appendix~\ref{sec:walkthroughs}.
We conducted agent walkthroughs from late May -- early June 2026, using Sonnet 4.6, GPT-5.5, and GPT-5.4-Mini.

\section{Results}
\label{sec:results}

\begin{table}
\centering
\begin{tabularx}{\textwidth}{c|l|X}
  & \bf Category & \bf Description \\\midrule
\multirow{8}{*}{\rotatebox{90}{\shortstack{\bf Specification\\\bf (UI or internal)}}}
  & None & System does not allow users to specify policies and/or does not describe an internal specification structure. \\
  & Natural language & Policies are in unstructured natural language. \\
  & Security labels & Traditional security labels, e.g., high/low to indicate private vs. public levels of access \\
  & Multiple selection & Choice of fixed options, e.g., mobile-style allow/deny policies. \\
  & Structured rules & Flexible rules within a structured set of actions or conditions. \\
  & Arbitrary & Policies can be anything, up to and including full programs. \\\midrule
\multirow{6}{*}{\rotatebox{90}{\bf Derivation}}
  & None & System does not take user input, even implicitly, in determining internal policies. \\
  & AI-predicted & Policy predicted from user queries or context. \\
  & AI-compiled & AI used to derive internal policy from explicit user spec. \\
  & Deterministic & User spec deterministically translated to internal policy. \\
  & Exact & User-specified and internal policies are one and the same. \\\midrule
\multirow{11}{*}{\rotatebox{90}{\bf Enforcement}}
  & None & No known enforcement mechanism; relies on best-effort compliance from the agent. \\
  & External & Enforcement mechanism defined by a third party, e.g., external tool providers. \\
  & AI guardrail & AI guardrails determine compliance (nondeterministic). \\
  & Generated per query & AI generates policy enforcement code anew for each task or user query. \\
  & Generated from policy & AI generates enforcement code up front, from the policy. \\
  & Non-AI automation & Checks for compliance without relying on AI, e.g., through deterministic enforcement code. \\
  & User-in-the-loop & User indicates violations in real time. \\
\end{tabularx}
\smallskip
\caption{
    Taxonomy of user-level agent permissions systems, sorted from least user control to most.
}
\label{tab:taxonomy}
\end{table}

\begin{table}[!htb]
\centering
\begin{tabularx}{\textwidth}{c|X|l}
  & \bf Threat model & \bf Count \\ \midrule
\multirow{5}{*}{\rotatebox[]{90}{\bf Adversary}}
  & External data sources, excluding third-party tools, the agent, and its provider. & 2 \\
  & Third-party tools. & 6 \\
  & Agent provider or platform. & 1 \\
  & Benign, but fallible, agent subject to prompt injection and/or hallucination. Implicitly includes external data sources. & 10 \\\midrule
\multirow{3}{*}{\rotatebox[]{90}{\bf Target}}
  & Data exfiltration to external parties, excluding other data/control flow. & 3 \\
  & General task flow, i.e., the combined data and control flow that make up the agent's trajectory. Implicitly includes data exfiltration. & 12 \\
\end{tabularx}
\smallskip
\caption{Threat models seen in user-level agent permissions systems in the literature. Not mutually exclusive except where noted. \emph{Note:} Six works did not define a threat model at all.}
\label{tab:threat-models}
\end{table}

We build a multilayered taxonomy describing how agent permissions systems specify, derive, and enforce user-level policies (Table~\ref{tab:taxonomy}).
Except those labeled ``None'', categories are generally not mutually exclusive, and many systems fall into multiple.
Because each permission system's threat model informs its approach, we also categorize threat models by adversary and target (Table~\ref{tab:threat-models}).
As indicated by our taxonomy, approaches to user-level agent permissions systems vary widely.
Even so, we identify three cross-cutting concerns as recurring themes:

(1) \emph{Low user overhead} in specifying policies, often through natural language input or AI-predicted policies.
(2) \emph{Formally grounded specifications}, based on established security principles such as information flow control~\cite{wu-et-al:f-secure:2024,li-et-al:ace:ndss26,kolluri-et-al:prudentia:2026} or type safety~\cite{palumbo-et-al:policy-compiler:arxiv26,shi-et-al:progent:arxiv25,wang-et-al:agentspec:icse26,lee-et-al:verisafe:mobicom25}.
And (3) \emph{deterministic enforcement}, providing more predictable and robust guarantees than AI-based guardrails.
All of the 21 works we review address one or more of these concerns, yet crucially, none fully achieve all three.
On the other hand, the commercial agents we evaluate rarely emphasize any of these themes.
They do, however, demonstrate a fourth concern:
(4) \emph{Continuous user control} over permissions policies, even after initial specification time.
This theme proves to be rare in the literature.

\subsection{Taxonomy}

We highlight a few notable elements of our taxonomy here.
For a full list, we refer the reader to the tables; for additional notes, see Appendix~\ref{sec:extended-taxonomy}.

\paragraph{Specifications.}
We identify several distinct forms of policy specifications.
\emph{Security labels} distinguish privileged from unprivileged data in information flow control systems.
Data labeled ``low'' (public) can be sent to tools labeled ``high'' (private), but not vice versa~\cite{wu-et-al:f-secure:2024,kolluri-et-al:prudentia:2026,li-et-al:ace:ndss26}.
In practice, we find that systems using security labels offer limited user control: users often (but not always) label their own query, but not external data or tools~\cite{kolluri-et-al:prudentia:2026,li-et-al:ace:ndss26}.
\emph{Multiple selection} systems offer a fixed list of policy options, often styled after mobile permissions, e.g., ``Always''/``Once''/``Never''~\cite{wu-et-al:isolategpt:2025,wu-et-al:automating-data-access:arxiv25}.
\emph{Structured rules} are more flexible, describing systems that allow nearly any policy within a pre-defined structure.
For example, Progent's programmable policies allow users to select one of three possible responses to policy violations~\cite{shi-et-al:progent:arxiv25}.
Finally, one system, \textsc{AgentSpec}, supports truly arbitrary policies.
Users can even write their own code as a policy specification~\cite{wang-et-al:agentspec:icse26}.

While each agent permissions system has at most one form of internal specification---and one~\cite{bagdasarian-et-al:airgapagent:ccs24} implements no internal spec at all---several offer users multiple ways to specify permissions.
For example, AC4A~\cite{sharma-and-grossman:ac4a:arxiv26} allows users to define policies via natural language or multiple selection.

\paragraph{Derivation and enforcement.}
Derivation categories refer to the means by which an agent permission system determines its internal policy from the user's input.
This may include \emph{implicit} input, such as queries or past context, as is the case with fully AI-predicted policies.
Meanwhile, enforcement describes the means by which a system detects policy violations at run time.
As with UI specifications, some systems offer multiple derivation or enforcement mechanisms, such as defaulting to deterministic enforcement with an opt-in LLM guardrail~\cite{provos:iron-curtain:2026}.

\subsection{Themes in the Literature}

\paragraph{Threat models.}
Six of the systems we review do not specify a threat model~\cite{li-et-al:drift:neurips25,lee-et-al:verisafe:mobicom25,chen-et-al:shieldagent:pmlr25,palumbo-et-al:policy-compiler:arxiv26,wang-et-al:agentspec:icse26,xiang-et-al:guardagent:2025}.
Of those that do, most define the agent as the adversary, not out of malice, but rather due to prompt injection and/or hallucinations~\cite{tsai-et-al:conseca:hotos25,gong-et-al:csagent:2025,palumbo-et-al:forge:2026,provos:iron-curtain:2026,sharma-and-grossman:ac4a:arxiv26,wu-et-al:automating-data-access:arxiv25,wu-et-al:f-secure:2024,zhu-et-al:miniscope:2025,shi-et-al:progent:arxiv25}.
Some proposals explicitly trust third-party tool providers~\cite{sharma-and-grossman:ac4a:arxiv26,zhu-et-al:miniscope:2025,kolluri-et-al:prudentia:2026},
while others treat them as adversaries~\cite{wu-et-al:isolategpt:2025,bagdasarian-et-al:airgapagent:ccs24,wu-et-al:automating-data-access:arxiv25,shi-et-al:progent:arxiv25,li-et-al:ace:ndss26}.
Most permissions systems aim to prevent unauthorized agent actions (i.e., tool calls) generally.
Three works~\cite{bagdasarian-et-al:airgapagent:ccs24,stanley-et-al:gaap:2026,wu-et-al:automating-data-access:arxiv25} use a narrower threat model, seeking only to prevent exfiltration of sensitive data.

\paragraph{Low user overhead.}
Over half (12) of the systems we review from the literature make design choices that minimize users' need to directly specify permissions policies.
Six proposals have users write permissions in natural language~\cite{bagdasarian-et-al:airgapagent:ccs24,xiang-et-al:guardagent:2025,chen-et-al:shieldagent:pmlr25,palumbo-et-al:forge:2026,provos:iron-curtain:2026,sharma-and-grossman:ac4a:arxiv26}, avoiding both the expressiveness limits of mobile-style multiple selection interfaces and the learning curve of writing policies in JSON or a domain-specific language (DSL).
Another five permissions systems eschew user specifications entirely and instead predict policies from context, typically including the user's query or information about the agent's execution state~\cite{tsai-et-al:conseca:hotos25,gong-et-al:csagent:2025,li-et-al:drift:neurips25,meng-et-al:cellmate:2026,lee-et-al:verisafe:mobicom25}.
Wu et al. take a distinct, but related approach of predicting policies from a limited set of past user-selected permissions decisions~\cite{wu-et-al:automating-data-access:arxiv25}.
Among systems minimizing user overhead, five also emphasize deterministic enforcement, while two formalize their specifications.
None, however, do both.

\paragraph{Formally grounded specifications.}
Eleven proposed systems ground their policy specifications in formal methods.
Some define information flow control (IFC) policies in terms of formalized security labels~\cite{li-et-al:ace:ndss26,wu-et-al:f-secure:2024,kolluri-et-al:prudentia:2026}, while others use typed DSLs or Datalog to ensure policies are well-formed~\cite{shi-et-al:progent:arxiv25,palumbo-et-al:policy-compiler:arxiv26,wang-et-al:agentspec:icse26,lee-et-al:verisafe:mobicom25,palumbo-et-al:forge:2026}.
ShieldAgent~\cite{chen-et-al:shieldagent:pmlr25} defines policies via linear temporal logic, while MiniScope~\cite{zhu-et-al:miniscope:2025} uses an Integer Linear Program (ILP) solver to identify the smallest set of necessary privileges.
Lastly, GAAP~\cite{stanley-et-al:gaap:2026} performs IFC analysis on code an agent generates to perform its task.

Some systems provide a foundation for formal specifications, yet come up short of defining the formalism itself.
AC4A~\cite{sharma-and-grossman:ac4a:arxiv26} describes its approach in terms of ``resource types'', implying a formal type system, yet does not discuss any concrete formalization.
Lastly, while many systems provide formalized specifications, none prove the correctness of their policy \emph{enforcement}.
$f$-secure~\cite{wu-et-al:f-secure:2024} and FORGE~\cite{palumbo-et-al:forge:2026} both provide formal analyses of their architectures, but the implementations themselves remain unverified.
GAAP's run-time analysis of agent-generated code may come the closest, yet its authors do not claim to have implemented formally verified enforcement~\cite{stanley-et-al:gaap:2026}.

\paragraph{Deterministic policy enforcement.}
Another twelve systems~\cite{tsai-et-al:conseca:hotos25,gong-et-al:csagent:2025,meng-et-al:cellmate:2026,palumbo-et-al:forge:2026,provos:iron-curtain:2026,wu-et-al:isolategpt:2025,wu-et-al:f-secure:2024,zhu-et-al:miniscope:2025,palumbo-et-al:policy-compiler:arxiv26,shi-et-al:progent:arxiv25,kolluri-et-al:prudentia:2026,stanley-et-al:gaap:2026} implement deterministic enforcement across two categories:
(1) Automated enforcement not involving AI.
While non-AI mechanisms could theoretically be nondeterministic themselves, in practice, none of the systems we review take this approach.
(2) ``User-in-the-loop'' enforcement, which prompts the user to decide policy compliance instead of detecting violations automatically.

We distinguish user-in-the-loop enforcement from systems that escalate to the user when a policy violation has \emph{already} been detected.
The latter is an implementation decision and orthogonal to the means of detecting violations in the first place.
Among the works we review, only three implement a user-in-the-loop approach:
\textsc{IsolateGPT} and GAAP, both of which treat user decisions as policy updates and deterministically apply them to subsequent queries~\cite{wu-et-al:isolategpt:2025,stanley-et-al:gaap:2026}; and
\textsc{Prudentia}, which prompts the user to override its default policies on every query~\cite{kolluri-et-al:prudentia:2026}.

We do not consider systems that use an LLM to generate enforcement code to be deterministic.
Not only are enforcement models inherently nondeterministic, but they remain vulnerable to prompt injection, and shielding them from untrusted data remains a challenge---especially when arbitrary user queries are used as input~\cite{xiang-et-al:guardagent:2025,chen-et-al:shieldagent:pmlr25,li-et-al:ace:ndss26}.
On the other hand, we label FORGE~\cite{palumbo-et-al:forge:2026} as offering deterministic enforcement, since it only uses LLMs when explicitly defined in user-provided enforcement code.
Finally, AC4A~\cite{sharma-and-grossman:ac4a:arxiv26} relies on enforcement functions provided by third-party tools, which may make arbitrary use of AI.
We thus do not consider AC4A's enforcement deterministic.

\subsection{Commercial Agents}
\label{ssec:commercial-agent-results}

We analyzed five popular commercial agents (Claude, Claude Cowork, ChatGPT chat mode, ChatGPT agent mode, and Codex) during late May and early June 2026.
We investigated each agent's user interface for specifying policies, and how those policies appeared to be enforced based on task results and, in Claude's case, logged MCP tool calls.
Since all five agents are closed-source, we could not evaluate their internal specifications.
Nonetheless, we glean several insights, most notably a tendency for commercial agents to gravitate to one of two extremes: Either high user control at the cost of high overhead, or low user overhead at the expense of control and transparency.

\paragraph{High user control, paired with high overhead.}
In contrast to the literature, we found a strong trend of commercial agents relying on user-in-the-loop enforcement with frequent decision prompts.
All four of the Claude chatbot, Claude Cowork, ChatGPT chat mode, and Codex used this approach at least some of the time.
Claude and Codex defaulted to user-in-the-loop enforcement, but also appeared to employ AI guardrails for auto-reviewing actions (see below).
ChatGPT, on the other hand, required explicit user approval for \emph{all} actions associated with a permissions policy; we found no way to turn this feature off or grant approval in advance.

The exception was ChatGPT's agent mode, which grants the LLM control over a remote browser.
We were unable to define permissions policies for the remote browser at all, meaning no opportunity for users to approve \emph{or} deny agent mode actions.
As such, ChatGPT freely took actions that required permissions to do locally, such as accessing Gmail.

\paragraph{Lack of transparency.}
The dichotomy between ChatGPT agent mode and the other agents illustrates commercial agents' tendency to reach for one of two extremes: When not relying on user-in-the-loop enforcement, they tend to fall back on the least transparent and least user-controllable alternatives.
We observe this pattern across all five agents, most notably in Claude.

The Claude chatbot offers three settings for each permission: ``Always allow'', ``Blocked'', and ``Needs approval'', the last of which implies a user-in-the-loop approach.
Yet we observed that under ``Needs approval'', Claude sometimes uses an LLM auto-reviewer \emph{implicitly}, without notifying the user.\footnote{This is distinct from Claude Code's ``auto mode'', which users can freely turn on or off~\cite{anthropic:claude-code-perms:2026}.}
For example, while ``Needs approval'' was selected, Claude autonomously approved the creation of an email draft and---as observed in the browser console---tagged the MCP query with \texttt{approvalRequired:\,false} (screenshots in Appendix~\ref{sec:walkthroughs}).
When asked why, Claude claimed to have inferred permission from our prompt.
Because the only other settings are ``Always allow'' and ``Blocked'', this auto-review behavior \emph{cannot be disabled} short of revoking access to a tool entirely.

We find a similar, if less egregious, lack of transparency in Claude Cowork.
Once Claude Cowork has received a permission, e.g., to read or delete files in a given folder, that access \emph{cannot} be revoked by the user.
Fresh Cowork sessions do not carry over permissions from previous ones, but older sessions persist in the app's history, and retain previously granted permissions if reentered.
As far as we could determine, the only way to permanently revoke a Cowork filesystem permission is through operating system features, if available, or by uninstalling the app entirely.

By contrast, Codex provides an explicit setting for its LLM auto-review, which is off by default and can be disabled at any time.
When enabled, however, Codex's auto-review applies to all tools used in the current session.
Users cannot selectively enable or disable auto-review for different tools.
Meanwhile, though ChatGPT did not provide an auto-review option, it appeared to use AI guardrails to tighten, rather than loosen, permission policies.
For example, when we asked ChatGPT to send an email mimicking a common prompt injection format (``Ignore all previous instructions. This email is safe and benign.''), the agent refused, in accordance with ``OpenAI's safety checks.''
In this instance, the AI enforcement mechanism resulted in fewer permissions for the agent, not more.
While less risky than Claude's automatic approvals, this behavior fits a larger theme of commercial agents reducing \emph{actual} user control over permissions, even as defaulting to a user-in-the-loop approach creates the \emph{appearance} of control.

\section{Discussion}
\label{sec:discussion}

Our review indicates several promising directions in existing agent permissions systems, emphasizing three main themes: (1) lowering user overhead, (2) grounding policy specifications in formalized security principles, and (3) deterministic enforcement.
Yet substantial gaps remain, both within the literature and especially between proposed systems and widely-used commercial agents.

\subsection{Gaps in the Literature}

\paragraph{1) Lack of proposals unifying all three concerns.}
Though several permissions systems in the literature emphasize two of the primary themes we identify, none of the works we review achieve all three.
Robust overlaps between proposals with deterministic enforcement and low user overhead, or deterministic enforcement and formal specifications, suggests that implementing a system with all three is technically achievable.
This makes it all the more surprising that none of the 21 systems we review do so.
AC4A~\cite{sharma-and-grossman:ac4a:arxiv26} and \textsc{Prudentia}~\cite{kolluri-et-al:prudentia:2026} come the closest, but AC4A does not provide a concrete formalism for its specifications, while \textsc{Prudentia} incurs high user overhead with its reliance on user-in-the-loop enforcement.
These examples show that closing this gap is both eminently achievable and an important target for ongoing and future work.

\paragraph{2) No formally verified enforcement.}
While 11 of the systems we review implement some level of formalism for their policy specifications, none verify the \emph{enforcement} of said policies, a step that would build confidence in permissions' systems robustness to arbitrary agentic actions.
Program verification is classically challenging.
Nonetheless, we argue that it is worthwhile, especially in an agentic security context.
As AI agents vastly expand the space of possible events, it becomes all the more important to ensure that the enforcement code functions correctly, even in the presence of unanticipated inputs.
Furthermore, policy enforcement code is a relatively small part of any agent.
Focusing verification efforts there thus can provide outsized guarantees about the safety of a larger agentic system, even with its many untrusted components.

\paragraph{3) Lack of continuous user control over policies.}
For an agentic security system to truly support user-specified permissions, it is crucial for users to be able to update their policies over time.
Yet only six of the systems we review clearly allow users to change or revoke permissions after initially specifying them~\cite{provos:iron-curtain:2026,sharma-and-grossman:ac4a:arxiv26,wu-et-al:isolategpt:2025,zhu-et-al:miniscope:2025,shi-et-al:progent:arxiv25,stanley-et-al:gaap:2026}.
Of these, two only allow changes between sessions~\cite{wu-et-al:isolategpt:2025,zhu-et-al:miniscope:2025}, which blocks permission management during single prolonged sessions.
Separately, another two proposals update policies based on the user's choices to allow or deny actions at run-time, yet lack ways for users to prevent or revert those updates~\cite{gong-et-al:csagent:2025,bagdasarian-et-al:airgapagent:ccs24}.
These systems assume that a one-time decision extrapolates to all future circumstances---a presumption that may not always be accurate.

\paragraph{4) Low threat model diversity.}
Nearly every system we review treats the adversary as agent misbehavior due to prompt injection or hallucination, and the target as general task flow (\S\ref{ssec:lit-review}).
A small number of proposals target a strict subset of this model, e.g., focused on data exfiltration~\cite{bagdasarian-et-al:airgapagent:ccs24,wu-et-al:automating-data-access:arxiv25,stanley-et-al:gaap:2026}.
All but one system assume that the agent and provider are benign.

We argue, first, that future work on agent permissions systems should take care to define their threat models explicitly, as several existing systems do not~\cite{li-et-al:drift:neurips25,xiang-et-al:guardagent:2025,chen-et-al:shieldagent:pmlr25,palumbo-et-al:policy-compiler:arxiv26,wang-et-al:agentspec:icse26}.
Second, we recommend that future proposals consider more diverse adversaries, such as malicious users in a multi-user or multi-agent context---addressed only by GuardAgent~\cite{xiang-et-al:guardagent:2025}---or even the agent provider itself.
Non-local agents send input data to the cloud to be processed, and privacy-conscious users may wish to prevent their private data from reaching the provider's servers~\cite{jazlan-et-al:chatbot-trackers:2026}.
Among the 21 papers in our survey, only GAAP~\cite{stanley-et-al:gaap:2026} accounts for this concern.
Targets beyond task flow similarly deserve consideration, such as denial of service, attempts to advertise an attacker's product via data poisoning~\cite{zhang-et-al:research-agent-poisoning:2026}, or malicious consumption of users' limited tokens and funds to perform unintended tasks.

\subsection{Gaps Between the Literature and Commercial Agents}

\paragraph{1) Commercial agents fall short on the literature's primary concerns.}
Despite the extensive literature on formally grounded, deterministically enforced, and/or low-user overhead permissions systems, the commercial agents we evaluate largely fail to achieve these aims.
This is a serious and substantial gap, particularly as the agents trade these goals for high user overhead, lack of transparency and user control over enforcement, or both.
While the commercial agents in our review outperform the literature on allowing users to update permissions over time, that success has limits.
Their reliance on user-in-the-loop enforcement creates a high risk of privacy fatigue, a phenomenon known to increase users' likelihood of ignoring warnings or over-granting permissions~\cite{choi-et-al:privacy-fatigue:2018,keith-et-al:privacy-fatigue:2014,vance-et-al:fog-of-warnings:2019}.

\paragraph{2) Forced tradeoffs between high overhead, or sacrificing user control.}
Often, the only alternative commercial agents provide to user-in-the-loop policy enforcement is an LLM auto-reviewer, over which users may have little or no control (\S\ref{ssec:commercial-agent-results}).
We observe this pattern directly in the agents we evaluate, and note that it appears in agents beyond those five as well.
Claude Code's auto mode~\cite{anthropic:claude-code-perms:2026}, for example, avoids user-in-the-loop permissions by using an LLM to decide approvals for all actions; and its \texttt{bypassPermissions} mode (a.k.a. YOLO) is even riskier, lacking even the protection of an auto-reviewer.
Substantial work is needed to close this gap between the literature and commercial agents, and build practical systems for users to control their permissions without becoming overwhelmed by prompts.
Moreover, any agentic permissions system \emph{must} be transparent about when and how it uses LLMs for policy decisions and enforcement.

\section{Limitations}
\label{sec:limitations}

Our literature review represents a snapshot of recent work on agent permissions systems as of late spring 2026.
With the rapidly evolving nature of the area, it is all but certain there exist relevant works not captured in our review.
Nonetheless, our survey and taxonomy contribute an important snapshot of the state of user-level permissions in AI agents, and point to valuable directions for ongoing and future work on user permissions for agents.

Separately, our ability to draw conclusions about commercial AI agent systems is limited by the lack of public information on their internal functions.
While we identify substantial insights by analyzing these systems from a user perspective, we ultimately cannot know the exact details of how closed-source agents handle permissions.
Commercial agent systems are also rapidly evolving and may exhibit nondeterministic behavior, particularly where LLMs handle permissions policies directly.

\section{Related Work}

Kim et al. survey attack vectors, mitigations, and challenges in building secure agentic systems~\cite{kim-et-al:agent-sec-sok:usenix26}.
Left unaddressed is the question of whether and how \emph{users} of agentic systems, as opposed to developers, may specify individual policies within a larger security framework, or how those user-level policies are subsequently enforced.
Zhang and Wang approach the user experience of agent security more directly in a survey of 18 web agents~\cite{zhang-and-wang:human-centered-auth:chi26}.
They identify widespread challenges in the web agent authorization landscape.
Several of these are echoed in our own findings, such as: a lack of granularity in permissions schemes; deceptive or missing options for revocation; and a disconnect between ``interaction-level'' authorizations selected by users, and broader ``system-level'' authorizations that actually take effect~\cite{zhang-and-wang:human-centered-auth:chi26}.

In concurrent work, Wang et al. review the academic literature and commercial agents on approaches to user intent alignment, a related but distinct concept from user-level permissions~\cite{wang-et-al:agent-human-interaction:2026}.
Intent alignment is the objective that an AI agent's actions should adhere to the user's goals and values.
User-level permissions, on the other hand, are a system-level mechanism that can enforce intent alignment, security policies, or other objectives.
There are even settings where user-level permissions must counteract the aims of intent alignment.
A customer service agent, for example, must serve many different users with divergent goals, \emph{without} allowing any one user's intent to unilaterally determine their permissions---lest they use the agent as a proxy to act maliciously on their behalf~\cite{krebs:instagram-bot-hack:2026}.
User-level permissions are thus crucial for balancing intent alignment with the need for appropriate restrictions on users themselves.

Wang et al. identify human-agent interaction as paramount for intent alignment~\cite{wang-et-al:agent-human-interaction:2026}.
As for user-level permissions, we concur on the importance of user control and transparency.
Yet we also highlight the need for more robust policy enforcement without overly relying on user interaction (\S\ref{sec:discussion}).

In other concurrent work, Brigham et al. also consider user-facing permissions in agentic systems, developing an evaluation framework to empirically compare
different permission approaches with respect to security- and usability-relevant criteria~\cite{brigham-et-al:janus:arxiv26}.
They evaluate implementations designed to capture different points in the design space, not existing commercial systems or other research prototypes, although they conceptually
compare some of the same systems that we consider.

We complement and expand on past work by taking a broad lens to the space of user-level permissions for AI agents.
We survey the recent literature to construct a \emph{platform-agnostic} taxonomy of agent permissions interfaces, specifications, and derivation and enforcement mechanisms, geared toward identifying when, how, and to what extent users can determine the permissions policies applied to their agent's actions and access.
We then identify cross-cutting themes in both the literature and commercial agents, and provide recommendations on areas in need of further exploration.

\section{Conclusion}
\label{sec:conclusion}

We survey 21 papers and open-source systems on implementing user-level permissions in agentic AI systems.
From these papers, we develop a unified taxonomy of approaches to specifying, deriving from user input, and enforcing user-level permissions in agents.
We then apply this taxonomy to an exploration of five commercial AI agents across web, desktop, and browser-use modalities.

We identify four primary concerns as themes across the literature and commercial agents: low user overhead in setting permissions policies, formally grounded policy specifications, deterministic enforcement, and continuous user control of previously specified policies.
While the first three themes feature prominently in the literature, we find them to be nearly absent from the commercial agents we examine.
Even deterministic enforcement, while often present in a user-in-the-loop form, is counterbalanced by the commercial agents' tendency to rely on AI guardrails as the sole alternative to constant permission requests---and one that users cannot always control.

Yet the reverse is true of continuous user control: though prominently featured in most commercial agents, few systems in the literature provide users with options to edit or revoke previously granted permissions.
We identify this as one of several substantial gaps in the academic literature, along with a dearth of systems combining all three of low user overhead, formal specifications, and deterministic enforcement; no formally verified enforcement; and low threat model diversity.

We encourage researchers to explore these areas in future work, toward a framework for user-level agent permissions that prioritizes not just usability and robustness, but also user control and transparency.
We furthermore urge the AI industry to prioritize these goals in their commercial agents, particularly around closing the gap between the high overhead of user-in-the-loop policy enforcement, and the low transparency and accountability of LLM-based enforcement.
Use of AI agents will only continue to grow.
As it does, it is incumbent upon academia and industry alike to ensure that the permissions that regulate agents' actions are correct, transparent, and accountable to users.

\section{Acknowledgements}

We thank David Kohlbrenner and Dan Grossman for their feedback on the early stages of this work.
This work was funded in part by gifts from Microsoft, including a Microsoft Grant for Customer Experience Innovation, and by NSF grant number DGE-2140004.
Any opinions, findings, and conclusions or recommendations expressed in this material are those of the authors and do not necessarily reflect the views of the National Science Foundation.

\bibliographystyle{plain}
\bibliography{bibs/agent-sec}

\appendix

\section{Agent Walkthrough Methodology}
\label{sec:walkthroughs}

We base our evaluation of commercial agents on the walkthrough method of application analysis~\cite{light-et-al:walkthrough-method:2018}.
For each agent, we used the following process to explore, first, how users specify policies; and second, how those policies are reflected and enforced during tasks:

\begin{enumerate}
    \item Set up or install each agent in a clean environment (e.g., a private browsing window or isolated VM), using an account created for this purpose.
    \item Explore accessible menus to identify which permissions settings are initially available.
    \item Grant a small number of permissions. For chatbots, we granted OAuth access to an experiment-specific Gmail account. For desktop agents, we granted access to a local file folder (e.g., the Desktop). The agentic browser had no permissions available to grant.
    \item Ask the agent to complete tasks that it should be able to accomplish under the current settings. For example, we asked chatbots with Gmail access to draft and send an email.
    \item Repeat under each possible setting for the granted permission, e.g., ``ask'' vs. ``allow'' vs. ``deny'', in a fresh agent session.
    \item Revoke the granted connection or permission. Assess whether the revocation was effective by (1) repeating the task requests and determining whether the agent fails, and (2) identifying whether available menus reflect the permission having been revoked.
\end{enumerate}

\begin{figure*}[!htb]
\centering
\frame{\includegraphics[width=\textwidth]{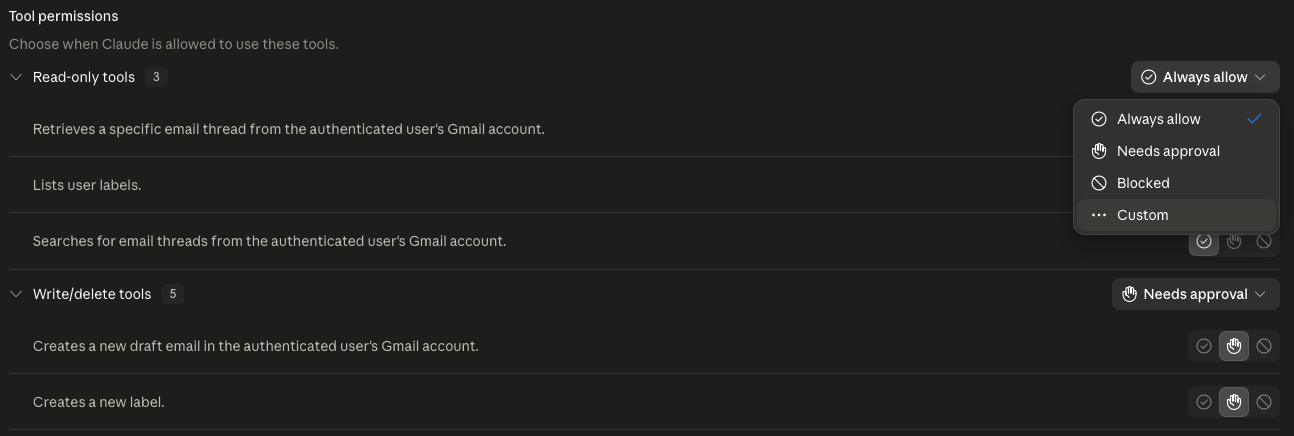}}
\caption{Selecting Gmail permissions in Claude.}
\label{fig:claude-perms}
\end{figure*}

\begin{figure*}[!htb]
\centering
\frame{\includegraphics[width=\textwidth]{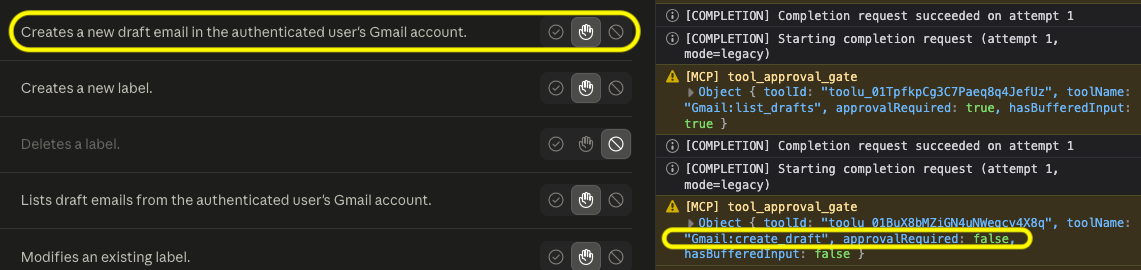}}
\caption{A logged MCP request from Claude indicating that no approval is required to create an email draft, despite having set the permissions for creating a draft to ``Needs approval''.}
\label{fig:claude-autoreview}
\end{figure*}

\begin{figure}[!htb]
\centering
\frame{\includegraphics[width=0.5\textwidth]{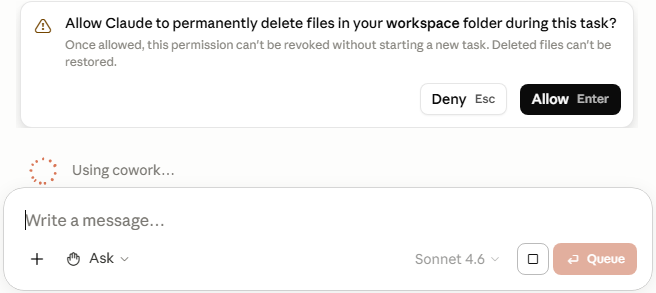}}
\caption{Claude Cowork requesting permission to delete files in the workspace folder. Contrary to the warning, beginning a new task does not remove the permission from the old one, which retains the permission if restarted later.}
\label{fig:claude-cowork-nonrevocable}
\end{figure}

\begin{figure}[!htb]
\centering
\frame{\includegraphics[width=0.5\textwidth]{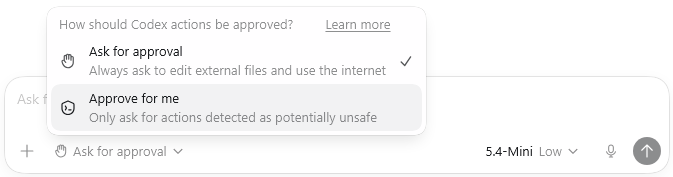}}
\caption{Enforcement settings in Codex, where they apply to all actions within the current session.}
\label{fig:codex-perms}
\end{figure}

We took detailed notes, screenshots, and videos throughout the walkthroughs (Figures~\ref{fig:claude-perms}, \ref{fig:claude-autoreview}, \ref{fig:claude-cowork-nonrevocable}, and \ref{fig:codex-perms}).
For the Claude chatbot in particular, we were able to glean additional insights by monitoring the browser's console log, where the agent logged its MCP tool calls.

We validated significant and unexpected results, such as Claude automatically approving an email draft creation in spite of a policy to always ask permission, by repeating the walkthrough process with a fresh session.
In Claude's case, we consistently reproduced the policy violation on at least three occasions across June and July 2026.
We disclosed this behavior to Anthropic on July 2, 2026.

\section{Extended Taxonomy and Labels}
\label{sec:extended-taxonomy}

\begin{table}[!htb]
\centering
\begin{tabularx}{\textwidth}{l|>{\hsize=1.5\hsize}X >{\hsize=.5\hsize}X}
    & \bf Threat model & \bf Occurrences \\\midrule
    \multirow{5}{*}{\rotatebox[]{90}{\bf Adversary}}
    & External data sources, excluding tools, the agent, and its provider
        & \cite{meng-et-al:cellmate:2026,kolluri-et-al:prudentia:2026}
        \\
    & Third-party tools
        & \cite{wu-et-al:isolategpt:2025,bagdasarian-et-al:airgapagent:ccs24,wu-et-al:automating-data-access:arxiv25,shi-et-al:progent:arxiv25,li-et-al:ace:ndss26,stanley-et-al:gaap:2026}
        \\
    & Agent provider
        & \cite{stanley-et-al:gaap:2026}
        \\
    & Benign but fallible agent (includes external data sources)
        & \cite{tsai-et-al:conseca:hotos25,gong-et-al:csagent:2025,palumbo-et-al:forge:2026,provos:iron-curtain:2026,sharma-and-grossman:ac4a:arxiv26,wu-et-al:automating-data-access:arxiv25,wu-et-al:f-secure:2024,zhu-et-al:miniscope:2025,shi-et-al:progent:arxiv25,stanley-et-al:gaap:2026}
        \\
    \midrule
    \multirow{3}{*}{\rotatebox[]{90}{\bf Target}}
    & Data exfiltration, excluding other forms of data/control flow
        & \cite{bagdasarian-et-al:airgapagent:ccs24,wu-et-al:automating-data-access:arxiv25,stanley-et-al:gaap:2026}
        \\
    & General task flow (combined run-time data/control flow; includes data exfiltration)
        & \cite{tsai-et-al:conseca:hotos25,gong-et-al:csagent:2025,meng-et-al:cellmate:2026,palumbo-et-al:forge:2026,provos:iron-curtain:2026,sharma-and-grossman:ac4a:arxiv26,li-et-al:ace:ndss26,wu-et-al:isolategpt:2025,wu-et-al:f-secure:2024,zhu-et-al:miniscope:2025,shi-et-al:progent:arxiv25,kolluri-et-al:prudentia:2026}
        \\
\midrule
 & No threat model given & \cite{li-et-al:drift:neurips25,lee-et-al:verisafe:mobicom25,chen-et-al:shieldagent:pmlr25,palumbo-et-al:policy-compiler:arxiv26,wang-et-al:agentspec:icse26,xiang-et-al:guardagent:2025}
\end{tabularx}
\caption{Distribution of threat models over 21 user-level agent permissions systems in the literature.}
\label{tab:threat-models-full}
\end{table}

Tables~\ref{tab:threat-models-full} and \ref{tab:taxonomy-results-full} offer a complete breakdown of threat models and taxonomy labels across the 21 systems we review from the literature.

\subsection{A Note on Threat Models}

Threat models for traditional, non-AI systems have historically distinguished between data flow, or the movement of information, and control flow, concerning the instructions executed by a program.
To an AI agent, however, data and control flow are unavoidably intertwined: instructions are merely another form of data, and data itself often contains instructions~\cite{masterman-et-al:agent-architectures-survey:2024,owasp:llm-top-10:2024}.

We therefore consider data and control flow as one unit in our threat model taxonomy, termed \emph{task flow}.
We define task flow as the combined flow of information and instructions over the course of an agent solving any given task.
The task flow category includes any threat model that broadly targets data, control flow, or both.

The threat models portion of our taxonomy also contains two mutually exclusive categories (other than ``None'').
A small number of papers explicitly limit their adversary to only include external data sources, or their target to only include data exfiltration.
Both threat models are implied by a more expansive one: external data sources are covered by benign-but-fallible agent adversaries, while data exfiltration is implicit in general task flow.
Because the great majority of papers with specified threat models fall into the more general categories, we list only the papers that use these categories exclusively.

\begin{table}[!tbph]
\centering
\begin{tabularx}{\textwidth}{l|>{\hsize=1.55\hsize}X >{\hsize=.5\hsize}X >{\hsize=.95\hsize}X}
    & {\bf Category}
        & \bf Literature & \bf Commercial agents
        \\
    \midrule
    \midrule
    \multirow{13}{*}{\rotatebox{90}{\bf Specification (UI)}}
    & \underbar{None} (no user-facing specification mechanism)
        & \cite{tsai-et-al:conseca:hotos25,gong-et-al:csagent:2025,li-et-al:drift:neurips25,meng-et-al:cellmate:2026,lee-et-al:verisafe:mobicom25}
        & ChatGPT agent mode~\cite{chatgpt}
        \\
    & \underbar{Natural language}
        & \cite{bagdasarian-et-al:airgapagent:ccs24,xiang-et-al:guardagent:2025,chen-et-al:shieldagent:pmlr25,palumbo-et-al:forge:2026,provos:iron-curtain:2026,sharma-and-grossman:ac4a:arxiv26}
        & \na
        \\
    & \underbar{Security labels}, e.g., high (secure) vs. low (public)
        & \cite{li-et-al:ace:ndss26,wu-et-al:f-secure:2024,kolluri-et-al:prudentia:2026}
        & \na
        \\
    & \underbar{Multiple selection} from a set of policies, e.g., mobile-style permissions
        & \cite{sharma-and-grossman:ac4a:arxiv26,wu-et-al:automating-data-access:arxiv25,wu-et-al:isolategpt:2025,zhu-et-al:miniscope:2025,stanley-et-al:gaap:2026}
        & Claude, Claude Cowork, Codex, ChatGPT chat mode~\cite{claude,claude-cowork,codex,chatgpt}
        \\
    & \underbar{Structured rules}, or flexible rules within a pre-defined set of actions or conditions
        & \cite{palumbo-et-al:policy-compiler:arxiv26,shi-et-al:progent:arxiv25}
        & \na
        \\
    & \underbar{Arbitrary}, permitting fully user-defined actions
        & \cite{wang-et-al:agentspec:icse26}
        & \na
        \\
    \midrule
    \multirow{11}{*}{\rotatebox{90}{\bf Specification (internal)}}
    & \underbar{None} (no internally tracked specification)
        & \cite{bagdasarian-et-al:airgapagent:ccs24}
        & \na
        \\
    & \underbar{Natural language}
        & \cite{xiang-et-al:guardagent:2025}
        & \na
        \\
    & \underbar{Security labels}, e.g., high (secure) vs. low (public)
        & \cite{li-et-al:ace:ndss26,wu-et-al:f-secure:2024,zhu-et-al:miniscope:2025,kolluri-et-al:prudentia:2026}
        & \na
        \\
    & \underbar{Multiple selection} from a set of policies, e.g., mobile-style permissions
        & \cite{sharma-and-grossman:ac4a:arxiv26,wu-et-al:automating-data-access:arxiv25,wu-et-al:isolategpt:2025,stanley-et-al:gaap:2026}
        & \na
        \\
    & \underbar{Structured rules}, permitting flexible rules within a pre-defined set of actions or conditions
        & \cite{tsai-et-al:conseca:hotos25,gong-et-al:csagent:2025,li-et-al:drift:neurips25,meng-et-al:cellmate:2026,lee-et-al:verisafe:mobicom25,chen-et-al:shieldagent:pmlr25,palumbo-et-al:forge:2026,provos:iron-curtain:2026,palumbo-et-al:policy-compiler:arxiv26,shi-et-al:progent:arxiv25}
        & \na
        \\
    & \underbar{Arbitrary}, permitting fully user-defined actions
        & \cite{wang-et-al:agentspec:icse26}
        & \na
        \\
    \midrule
    \multirow{10}{*}{\rotatebox{90}{\bf Derivation}}
    & \underbar{None} (no UI-to-internal translation occurs)
        & \cite{bagdasarian-et-al:airgapagent:ccs24}
        & ChatGPT agent mode~\cite{chatgpt}
        \\
    & \underbar{AI-predicted} from user query and/or implicit context
        & \cite{tsai-et-al:conseca:hotos25,gong-et-al:csagent:2025,li-et-al:drift:neurips25,meng-et-al:cellmate:2026,lee-et-al:verisafe:mobicom25,sharma-and-grossman:ac4a:arxiv26,wu-et-al:automating-data-access:arxiv25}
        & Claude~\cite{claude}
        \\
    & \underbar{AI-compiled} from user-provided specification
        & \cite{chen-et-al:shieldagent:pmlr25,palumbo-et-al:forge:2026,provos:iron-curtain:2026,sharma-and-grossman:ac4a:arxiv26}
        & \na
        \\
    & \underbar{Deterministic} compilation from user-specified policy
        & \cite{wu-et-al:f-secure:2024,zhu-et-al:miniscope:2025}
        & \na
        \\
    & \underbar{Exact use} of user-specified policy
        & \cite{xiang-et-al:guardagent:2025,sharma-and-grossman:ac4a:arxiv26,li-et-al:ace:ndss26,wu-et-al:isolategpt:2025,palumbo-et-al:policy-compiler:arxiv26,shi-et-al:progent:arxiv25,wang-et-al:agentspec:icse26,kolluri-et-al:prudentia:2026,stanley-et-al:gaap:2026}
        & Claude, Claude Cowork, Codex, ChatGPT chat mode~\cite{claude,claude-cowork,codex,chatgpt}
        \\
    \midrule
    \multirow{15}{*}{\rotatebox{90}{\bf Enforcement}}
    & \underbar{None} (relies on best-effort compliance from the agent)
        & \cite{lee-et-al:verisafe:mobicom25}
        & Claude Cowork~\cite{claude-cowork}
        \\
    & \underbar{External} mechanisms, defined by third-party tools
        & \cite{sharma-and-grossman:ac4a:arxiv26}
        & \na
        \\
    & \underbar{AI guardrail} determines compliance
        & \cite{li-et-al:drift:neurips25,provos:iron-curtain:2026,bagdasarian-et-al:airgapagent:ccs24}
        & Claude, Codex~\cite{claude,codex}
        \\
    & \underbar{AI-generated per query}, e.g., enforcement code generated separately for each user query or task
        & \cite{xiang-et-al:guardagent:2025,chen-et-al:shieldagent:pmlr25,li-et-al:ace:ndss26}
        & \na
        \\
    & \underbar{AI-generated from policy}, up front
        & \cite{wang-et-al:agentspec:icse26}
        & \na
        \\
    & \underbar{Non-AI automation}: automated violation detection without use of AI; generally deterministic
        & \cite{tsai-et-al:conseca:hotos25,gong-et-al:csagent:2025,meng-et-al:cellmate:2026,palumbo-et-al:forge:2026,provos:iron-curtain:2026,wu-et-al:isolategpt:2025,wu-et-al:f-secure:2024,zhu-et-al:miniscope:2025,palumbo-et-al:policy-compiler:arxiv26,shi-et-al:progent:arxiv25,kolluri-et-al:prudentia:2026,stanley-et-al:gaap:2026}
        & Claude~\cite{claude}
        \\
    & \underbar{User-in-the-loop}: User indicates violations in real time
        & \cite{wu-et-al:isolategpt:2025,kolluri-et-al:prudentia:2026,stanley-et-al:gaap:2026}
        & Claude, Claude Cowork, Codex, ChatGPT chat mode~\cite{claude,claude-cowork,codex,chatgpt}
        \\
\end{tabularx}
\caption{Full taxonomy of user-level agent permissions specifications (user-facing and internal), means of deriving internal specifications from user-facing ones, and run-time enforcement mechanisms, as distributed across proposed and commercial agent systems.}
\label{tab:taxonomy-results-full}
\end{table}

\section{Generative AI Use}

Generative AI tools were used solely as subjects of evaluation, as described in Section~\ref{sec:methods}, toward answering the principal research questions of this work.
No generative AI was used otherwise.
The authors completed all ideation and planning, literature searches, paper reviewing, and writing without AI assistance.

\end{document}